\documentclass[nofootinbib,superscriptaddress,eqsecnum,tightenlines,11pt]{revtex4}

\usepackage{hyperref}
\usepackage{graphicx}
\usepackage{amsmath,amssymb,amsfonts,amsthm,latexsym,stmaryrd}

\def\be{\begin{equation}}
\def\ee{\end{equation}}   
\def\ba{\begin{eqnarray}}
\def\ea{\end{eqnarray}}
\def\bas{\begin{subequations}\begin{eqnarray}}
\def\eas{\end{eqnarray}\end{subequations}}

\def\lp{\ell_\text{Pl}}

\def\de{\mathrm{d}}

\def\f{\frac}

\def\SU{\text{SU}}

\def\nn{\nonumber}

\begin{document}

\title{Black holes as gases of punctures with a chemical potential:\\
Bose--Einstein condensation and logarithmic corrections to the entropy}

\author{Olivier Asin}
\affiliation{Laboratoire de Math\'ematiques et Physique Th\'eorique, CNRS (UMR 7350), F\'ed\'eration Denis Poisson,Universit\'e Fran\c{c}ois Rabelais, Parc de Grandmont, 37200 Tours, France}

\author{Jibril Ben Achour}
\affiliation{Laboratoire APC -- Astroparticule et Cosmologie, Universit\'e Paris Diderot Paris 7, 75013 Paris, France}

\author{Marc Geiller}
\affiliation{Institute for Gravitation and the Cosmos \& Physics Department, Penn State, University Park, PA 16802, U.S.A}

\author{Karim Noui}
\affiliation{Laboratoire de Math\'ematiques et Physique Th\'eorique, CNRS (UMR 7350), F\'ed\'eration Denis Poisson,Universit\'e Fran\c{c}ois Rabelais, Parc de Grandmont, 37200 Tours, France}
\affiliation{Laboratoire APC -- Astroparticule et Cosmologie, Universit\'e Paris Diderot Paris 7, 75013 Paris, France}

\author{Alejandro Perez}
\affiliation{Aix Marseille Universit\'e, CPT, CNRS (UMR 7332), 13288 Marseille, France\\ and Universit\'e de Toulon, CPT, CNRS (UMR 7332), 83957 La Garde, France}

\begin{abstract}
We study the thermodynamical properties of black holes when described as gases of indistinguishable punctures with a chemical potential. In this picture, which arises from loop quantum gravity, the black hole microstates are defined by finite families of half-integers spins coloring the punctures, and the near-horizon energy measured by quasi-local stationary observers defines the various thermodynamical ensembles. The punctures carry excitations of quantum geometry in the form of quanta of area, and the total horizon area $a_\text{H}$ is given by the sum of these microscopic contributions. We assume here that the system satisfies the Bose--Einstein statistics, and that each microstate is degenerate with a holographic degeneracy given by $\exp\big(\lambda a_\text{H}/\ell_\text{Pl}^2\big)$ and $\lambda>0$. We analyze in detail the thermodynamical properties resulting from these inputs, and in particular compute the grand canonical entropy. We explain why the requirements that the temperature be fixed to the Unruh temperature and that the chemical potential vanishes do not specify completely the semi-classical regime of large horizon area, and classify in turn what the various regimes can be. When the degeneracy saturates the holographic bound ($\lambda=1/4$), there exists a semi-classical regime in which the subleading corrections to the entropy are logarithmic. Furthermore, this regime corresponds to a Bose--Einstein condensation, in the sense that it is dominated by punctures carrying the minimal (or ground state) spin value $1/2$.
\end{abstract}

\maketitle

\section{Introduction}

\noindent One of the notable achievements of loop quantum gravity (LQG hereafter) \cite{LQG1,LQG2,LQG3} is its ability to describe the quantum geometry of black holes and the nature of the microstates contributing to the Bekenstein--Hawking entropy \cite{BHentropyLQG1,BHentropyLQG2,BHentropyLQG3,BHentropyLQG4,BHentropyLQG5,BHentropyLQG6,BHentropyLQG7}. This computation relies on the quasi-local notion of isolated horizons \cite{IH}, which has been introduced as a way to describe stationary spacetimes while still allowing for exterior dynamical processes to happen, and has the advantage of also accounting for cosmological horizons. In this framework, the computation of the entropy relies essentially on the idea that the macroscopic horizon area is realized as the sum of microscopic contributions (quanta of area) carried by spin network edges puncturing the horizon, and the counting of the microstates leads to the Bekenstein--Hawking area law provided that a free parameter of the theory, the Barbero--Immirzi parameter, is fixed to a particular finite and real value.

However, although this description has the advantage of being background-independent and mathematically rigorous, it makes quite hard the contact with other existing semi-classical results on black hole kinematics and dynamics. For this reason, a complementary approach has been recently developed, based on how near-horizon stationary observers would describe a black hole space-time within LQG \cite{ale-amit,Frodden:2011eb}, and whose setup is essentially as follows. First, one considers observers at a fixed distance $\ell\ll\sqrt{a_\text{H}/(4\pi)}$ from the horizon, with four-velocity $u^\mu=\xi^\mu/\sqrt{\xi\cdot\xi}$, where $\xi^\mu$ is the Killing vector corresponding to the null normal to the horizon. Then, with a careful treatment of the boundary terms in the Hamiltonian framework, one can show that there is a well-defined notion of near-horizon energy measured by these observers. This latter is simply given by
\be
E=\f{a_\text{H}}{8\pi\ell},
\ee
where the horizon area $a_\text{H}$ can be written with the usual area operator of LQG, and where $1/\ell$ is the acceleration (of $u^\mu$) needed in order for the observers to remain at rest. Finally, because of this acceleration and of the local correspondence between the Rindler geometry and that of the black hole spacetime, one assumes that the observers measure a near-horizon temperature given by the Unruh temperature $T_\text{U}=\lp^2/(2\pi\ell)$ \cite{UnruhT}.

With these inputs, one arrives at a statistical mechanical picture in which the near-horizon observers describe the black hole as a gas of punctures whose thermodynamical properties are encoded in a grand canonical partition function $\mathcal{Z}_\text{GC}(\beta,\mu)$, where $\mu$ is the chemical potential conjugated to the number $n$ of punctures and $\beta$ is the inverse temperature. Because of the usual equivalence between thermodynamical ensembles, one can choose to work in the canonical ensemble with partition function $\mathcal{Z}_\text{C}(\beta)$, which requires to fix the chemical potential to $\mu=-T({\partial S_\text{GC}}/\partial n)|_E$, where $S_\text{GC}$ is the entropy in the grand canonical ensemble and $T=\beta^{-1}$ is the temperature. Since the area operator (or rather its eigenvalues) of LQG entering the definition of the near-horizon energy $E$ is a function of the Barbero--Immirzi parameter $\gamma$, this equation can be used to determine the chemical potential as a function $\mu(\gamma)$ of $\gamma$. Then, one can decide on a value of the chemical potential, which in turn determines a value for $\gamma$. In this work we are not going to focus or to comment on the precise fixation of the Barbero--Immirzi parameter, but rather study the role of the chemical potential itself.

In fact, the appearance of an a priori non-vanishing chemical potential leads to both conceptual and technical questions. On the one hand, one can argue that, since the level spacing between successive area eigenvalues vanishes  for large eigenvalues (and therefore for large horizon areas), the energy required to create or annihilate a puncture becomes arbitrarily small for large black holes, which in turn implies that the chemical potential can be set to zero. A consequence of this choice is that the number $n$ of punctures is not an observable (i.e. the punctures behave in some sense like photons). As such, this picture is more easily compatible with the usual semi-classical understanding that we have of black holes. On the other hand, one can assume that the chemical potential is not vanishing, which implies that the number $n$ of punctures becomes an observable, just like the number of molecules for a gas in a box. This would seem to indicate that, in addition to the usual macroscopic charges (the mass $M$, the electric charge $Q$, and the angular momentum $J$), a quantum notion of black hole hair, or global charge $n$, arises from the underlying LQG description. It is not clear at present which point of view should be adopted, and what the full consistent picture is. Furthermore, besides this question about the role of the chemical potential, the description of a quantum black hole as a gas of punctures depends on additional inputs that are not yet derived from first principles. In particular, one has to decide on a choice of statistics for the punctures, as well as on a degeneracy for the microstates.

Because of this freedom in choosing the various parameters defining the statistical physics description (i.e. the partition function), it is important to study in a systematic manner the different possibilities and see if they are compatible with our semi-classical understanding of the thermodynamical properties of black holes. This investigation was started in \cite{Ghosh:2013iwa}, where it was shown that the grand canonical entropy for large black holes reproduces the Bekenstein--Hawking formula under the assumptions that i) the chemical potential is vanishing, ii) the punctures are indistinguishable (with either a bosonic or fermionic statistics), iii) the degeneracy of the punctures is exponential in the area, i.e. of the form $\exp\big(\lambda a_\text{H}/\lp^2\big)$ with $\lambda>0$, as justified from the expected QFT behavior of entanglement entropy \cite{Ghosh:2013iwa,holo}. Furthermore, with these assumptions, it was shown that the number of punctures scales as $n\propto\sqrt{a_\text{H}}/\lp$, that the system is dominated by large spin contributions, and that the holographic bound is satisfied in the sense that $\lambda=1/4$. However, the subleading corrections to the entropy that follow from these inputs are of the form $\sqrt{a_\text{H}}/\lp$, which is larger than the corrections in $\log\big(a_\text{H}/\lp^2\big)$ appearing in other treatments \cite{log1,log2,log3}. The present paper is partly devoted to the study of these subleading corrections, and of the conditions under which they can be logarithmic.

Even if the case $\mu=0$ is appealing because it eliminates the need for explaining the physical meaning of the new black hole charge $n$, here we would like to study the consequences of having (at least to start with) an a priori non-trivial chemical potential and a given quantum statistics for the punctures. Furthermore, we would like to understand whether phenomena such as condensation can actually be realized for a gas of bosonic punctures. We are in fact going to show that Bose--Einstein condensation can indeed be achieved (under suitable conditions), in the sense that there exists a phase in which the system is dominated by punctures carrying spin 1/2, and that in this phase the subleading corrections to the entropy scale as $\log\big(a_\text{H}/\lp^2\big)$.

\subsection*{Outline}

\noindent In this paper we consider the description of black holes as gases of punctures with a chemical potential, and study the thermodynamical properties of such statistical mechanical models. The black hole microstates are defined by (almost empty) families of integers $\vec{n}_j$ such that the horizon area is $a_\text{H}=\sum_jn_ja_j$, where the expression for the quanta of area $a_j$ is recalled in \eqref{area spectrum}. We assume that the degeneracy of the microstates is holographic in the sense of equation \eqref{degeneracy} below. Our study is organized as follows.

In section \ref{sec:2}, we assume that the punctures satisfy the Maxwell--Boltzmann statistics, and compute the grand canonical partition function and related physical quantities such as the entropy, the mean energy, and the mean number of punctures. We show that with a fine tuning of the chemical potential it is possible to obtain logarithmic subleading corrections to the entropy, and that the corresponding value of the chemical potential depends on whether the degeneracy saturates the holographic bound or not.

In section \ref{sec:3}, we consider the Bose--Einstein statistics for the punctures, and perform a complete analysis of the thermodynamical properties of the system. We start by assuming that the holographic bound is saturated and that the temperature approaches the Unruh temperature at the semi-classical limit. In this case, we show that there exists a regime in which the subleading corrections to the entropy are logarithmic, and that this regime cannot be attained if the chemical potential is assumed to vanish from the onset. Furthermore, in this regime one observes a condensation, i.e. a domination of the number of punctures carrying a spin $1/2$. If we relax the condition that the holographic bound be saturated, and assume that the black hole temperature is fixed to the Unruh temperature, then the quantum corrections of the entropy can no longer be logarithmic. We conclude with a discussion in section \ref{sec:4}.

\subsection*{Some notations}

\noindent Throughout this paper, we will use the symbol $\simeq$ to denote an equality between two functions of variables $\delta_1,\ldots,\delta_k$, which holds up to subleading terms when $\delta_i\ll1$ (with possible conditions on the variables $\delta_i$), i.e.
\be
f\simeq g\qquad\text{iff}\qquad\lim_{\delta_1,\ldots,\delta_k\to0} \f{f(\delta_1,\ldots,\delta_k)}{g(\delta_1,\ldots,\delta_k)}=1.
\ee
Furthermore, we will use the symbol $f\rightarrow g$ to denote the fact that $f$ tends to $g$ in the semi-classical limit. In the core of the article, this arrow will only be used to denote a behavior at the semi-classical limit.

\section{Black holes as gases of punctures}
\label{sec:2}

\noindent In this section, we briefly review some aspects of the description of black holes as gases of punctures. For this, let us first focus on the microcanonical ensemble. This latter is defined by an energy $E=a_\text{H}/(8\pi\ell)$ (which is the near-horizon energy measured by a stationary observer at distance $\ell$ from the horizon \cite{Frodden:2011eb}) and a number $n$ of punctures. These punctures, which we label by integers $e\in\llbracket1,n\rrbracket$, can be thought of as being inherited from bulk LQG spin network excitations crossing the horizon of the black hole, and each carry a quantum of area taking the discrete values
\be\label{area spectrum}
a_e=8\pi\gamma\lp^2\sqrt{j_e(j_e+1)}.
\ee
Here $\lp$ is the Planck length, $\gamma\in\mathbb{R}$ is the Barbero--Immirzi parameter, and the labels $j_e$ are $\SU(2)$ half-integer spins. Heuristically, the punctures can be interpreted as (locally) non-interacting particles living on the horizon of the black hole.

\subsection{Degeneracy of the microstates}

\noindent Let us now consider a labeling of all the punctures by a set of spins $\vec{\jmath}_e=(j_1,\dots,j_n)$. We call this assignment of spins a ``quantum configuration'' of the black hole. The key quantity for the computation of the microcanonical entropy and for the generalization to other thermodynamical ensembles is the degeneracy $D(\vec{\jmath}_e)$ of states corresponding to a fixed black hole configuration (i.e. with fixed number of punctures $p$ and fixed set of spins $\vec{\jmath}_e$). Typically, this degeneracy is given by the dimension of the Hilbert space of Chern--Simons theory on a sphere with $n$ punctures. Explicitly, the formula for this dimension is
\be
D(\vec{\jmath}_e)=\f{2}{k+2}\sum_{d=1}^{k+1}\sin\left(\f{\pi d}{k+2}\right)^{2-n}\prod_{e=1}^n\sin\left(\f{\pi dd_e}{k+2}\right),
\ee
where $k$ is the level of the Chern--Simons theory (which is proportional to the horizon area $a_\text{H}$ and which can be sent to infinity at the semi-classical limit), 
$d_e=2j_e+1$ is the classical dimension of the spin $j_e$ representation space, and where the spins are restricted to half-integers values between $1/2$ and $k/2$. Starting from this dimension, we can obtain the total number of states for a black hole with $n$ punctures by summing over all the spin labels. This defines the quantity
\be
N(n,k)=\sum_{\vec{\jmath}_e}D(\vec{\jmath}_e),
\ee
where the sum runs over all possible spin values at each puncture, and therefore treats the punctures as being distinguishable.

Now, if one is interested in computing the microcanonical entropy by using the method of most probable configuration, it is convenient to switch from the representation in terms of spins labels to the representation in terms of spin occupation numbers. Using the multinomial formula, one can show that the above number of states can be written in the form
\be\label{sumn}
N(n,k)=\sum_{\vec{n}_j}\f{n!}{\prod_{j=1/2}^{k/2}n_j!}D(\vec{n}_j),
\ee
where
\be\label{dimn}
D(\vec{n}_j)=\f{2}{k+2}\sum_{d=1}^{k+1}\sin\left(\f{\pi d}{k+2}\right)^{2-n}\prod_{j=1/2}^{k/2}\sin\left(\f{\pi dd_j}{k+2}\right)^{n_j}.
\ee
In this new (but equivalent) expression, the occupation number $n_j$ represents the number of punctures carrying a spin label $j$. In the limit of large level $k$, the degeneracy \eqref{dimn} can be viewed as the Riemann sum of an integral over an angle (which gives the well-known formula for the number of classical $\SU(2)$ invariant tensors), and when the spins are large it is easy to show that its leading order term tends to $\log D(\vec{n}_j)\simeq\sum_jn_j\log(2j+1)$, 
where the sum is taken over all possible half-integers. More precisely, one can show that
\be
D(\vec{n}_j)\simeq P(\vec{n}_j)\prod_j (2j+1)^{n_j},
\ee
where $P(\vec{n}_j)$ is a rational function of the occupation numbers $n_j$ (and does not include an exponential function). It is this expression for the degeneracy (with $P(\vec{n}_j)=1$) which was taken in \cite{ale-amit} as the starting point for the study of the statistical mechanics of the quantum black holes in various thermodynamical ensembles. In a series of papers \cite{FGNP,BTZ,Achour:2014eqa}, it was shown that an analytic continuation of the Barbero--Immirzi parameter to $\gamma=\pm i$ leads to a holographic behavior for \eqref{dimn}.

In the present work, at the difference with \cite{ale-amit} and following \cite{Ghosh:2013iwa}, we are going to assume that the punctures are fundamentally indistinguishable, and that the degeneracy factor $D(\vec{n}_j)$ is holographic but does not necessarily saturates a priori the holographic bound. In the following subsection we briefly review the interesting physical properties that arise from these assumptions.

\subsection{Setup and results}

\noindent Let us consider an almost-vanishing family of integers $\vec{n}_j=(n_{1/2},n_1,\dots)$ labeling a quantum microstate of the black hole. This set of occupation numbers is such that the number of punctures is given by $n=\sum_jn_j$, while the area of the horizon is $a_\text{H}=\sum_jn_ja_j$. Furthermore, we assume that the quantum state is degenerate, with a degeneracy given by a holographic law. Following the notations of \cite{Ghosh:2013iwa}, we write this degeneracy as
\be\label{degeneracy}
D(\vec{n}_j)=\exp\left(\lambda\f{a_\text{H}}{\lp^2}\right)=\prod_j\exp\left(2\pi\gamma(1-\delta_\text{h})n_j\sqrt{j(j+1)}\right),\qquad\text{with}\qquad\lambda=\f{1}{4}(1-\delta_\text{h}).
\ee
The free parameter $\delta_\text{h}$ measures the failure to saturate the holographic bound, which is obtained for $\delta_\text{h}=0$.

The study of this statistical system has been carried out in details in \cite{Ghosh:2013iwa} under the assumptions that the punctures have no chemical potential $\mu$. Hence, the fugacity $z=\exp(\beta\mu)$, where $\beta$ is the inverse temperature, was fixed to $z=1$. Furthermore, the area spectrum \eqref{area spectrum} was replaced by the linear expression $a_e=8\pi\gamma\lp^2(j_e+1/2)$, which is only valid in the limit of large spins (this approximation is however self-consistent with the fact the system was shown to be dominated by large spins in the semi-classical limit).

Based on these assumptions, the grand canonical partition function and related thermodynamical quantities (such as the mean energy, the mean number of particles, and the specific heat) have been computed in \cite{Ghosh:2013iwa} at the semi-classical limit for the Maxwell--Boltzmann, the Bose--Einstein, and the Fermi--Dirac statistics. For these three choices of statistics, it can be shown that:
\begin{enumerate}
\item The requirement that the semi-classical limit occurs at the Unruh temperature $T_\text{U}$ for the local observers, which corresponds to the inverse temperature $\beta_\text{U}=1/T_\text{U}={2\pi\ell}/{\lp^2}$, imposes that the the holographic degeneracy \eqref{degeneracy} be saturated, i.e. that $\lambda-1/4\ll1$ or equivalently that $\delta_\text{h}\ll1$.
\item The large spins dominate at the semi-classical limit, in the sense that the mean spin of the punctures tends to infinity when $a_\text{H}$ becomes large.
\item The entropy of the gas of punctures reproduces the Bekenstein--Hawking result $S={a_\text{H}}/{4\lp^2}+S_\text{cor}$ with the correct factor of $1/4$, while the subleading corrections are of the form $S_\text{cor}\propto\sqrt{a_\text{H}}$ (the precise value of the multiplicative factor is not important and depends on the statistics). This large entropy is mainly due to the fluctuations of the number of punctures at the semi-classical limit.
\end{enumerate}

These results are physically very encouraging, a part from the subleasing corrections $S_\text{cor}$ to the entropy. Indeed, these corrections are typically expected to be logarithmic (as functions of $a_\text{H}$), and therefore one should look for a mechanism able to suppress the too large $\sqrt{a_\text{H}}$ contributions. It has been shown in \cite{Achour:2014eqa} that the introduction of a non-vanishing chemical potential allows to eliminate these too large corrections and to replace them by logarithmic ones, at least in the case where the (undistinguishable) punctures satisfy the classical Maxwell--Boltzmann statistics and when the holographic degeneracy is saturated (i.e. $\delta_\text{h}=0$). To understand how this comes about, let us briefly reproduce the main steps of the calculation in \cite{Achour:2014eqa} when $\delta_\text{h}\neq0$ (for the sake of generality).

\subsection{Grand canonical partition function in Maxwell--Boltzmann statistics}

\noindent When the punctures are taken to be undistinguishable, with a non-vanishing chemical potential $\mu$, with a Maxwell--Boltzmann statistic, and with a degeneracy \eqref{degeneracy}, the grand canonical partition function for the gas of punctures is given by
\be
\mathcal{Z}_\text{M}(\beta,\mu)=\sum_{\vec{n}_j}\f{D(\vec{n}_j)}{\prod_jn_j!}z^n\exp(-\beta E).
\ee
Here the sum runs over the set of almost-vanishing families $\vec{n}_j$ of integers, $E$ is the quasi-local energy of the gas, $n=\sum_jn_j$ represents the number of punctures, $D(\vec{n}_j)$ is the degeneracy \eqref{degeneracy}, and $z=\exp(\beta\mu)$ is the fugacity. The partition function can be written in the simpler form
\be\label{partition function maxwell}
\mathcal{Z}_\text{M}(\beta,\mu)=\exp\big(z\mathcal{Q}(\beta)\big),\qquad\text{with}\qquad\mathcal{Q}(\beta)=\sum_j\exp\big((\beta_\text{U}-\delta_\text{h}\beta_\text{U}-\beta)E_j\big),\qquad\text{and}\qquad E_j=\f{a_j}{8\pi l}.
\ee
Using the notations of \cite{Ghosh:2013iwa}, where $\delta_\beta=\beta/\beta_\text{U}-1$ and $\delta=\delta_\text{h}+\delta_\beta$, we can further write that
\be
\mathcal{Q}(\beta)=\sum_j\exp\left(-\delta\f{a_j}{4\lp^2}\right).
\ee
This expressions shows immediately that $\mathcal{Q}(\beta)$, and therefore $\mathcal{Z}_\text{M}(\beta,\mu)$, are only defined for $\delta>0$.

Thermodynamical quantities, such as the mean energy $\bar{E}$ and the mean number of punctures $\bar{n}$, are immediately obtained from the derivatives of $\mathcal{Z}_\text{M}(\beta,\mu)$ as
\be
\bar{E}=-\f{\partial}{\partial\beta}\log\mathcal{Z}_\text{M}(\beta,\mu)=z\sum_jE_j\exp\left(-\delta\f{a_j}{4\lp^2}\right),\qquad\text{and}\qquad
\bar{n}=z\f{\partial}{\partial z}\log\mathcal{Z}_\text{M}(\beta,\mu)=z\mathcal{Q}(\beta).
\ee
The mean horizon area $\bar{a}_\text{H}=8\pi l\bar{E}$ follows directly from this.

\subsection{Holography and entropy}

\noindent Following \cite{Ghosh:2013iwa}, let us set the temperature to be the Unruh temperature at the semi-classical limit, which schematically means that $\delta_\beta\rightarrow0$ when $\lp\rightarrow0$. More let us define the semi-classical limit by $\bar{a}_\text{H}\gg\lp^2$ and $\delta_\beta=0$. Note that the condition $\delta_\beta=0$ is only possible when $\delta_\text{h}\neq0$, which is what we are going to assume for the time being. We are going to show that the black hole degeneracy necessarily saturates the holographic bound (i.e. $\delta_\text{h}\rightarrow0$) at the semi-classical limit. More precisely, we can prove that the mean horizon area becomes macroscopic (with respect to $\lp^2$) when $\beta=\beta_\text{U}=T_\text{U}^{-1}$ only in the regime where $\delta_\text{h}\ll1$. In order to do so, we can write the mean energy in the form
\be
\bar{E}=z\pi\gamma T_\text{U}\sum_{k=1}^\infty\sqrt{k(k+2)}\exp\left(-\pi\gamma\delta\sqrt{k(k+2)}\right),
\ee
and use the inequalities
\be
z\pi\gamma T_\text{U}\sum_{k=1}^\infty k\exp\big(-\pi\gamma\delta(k+1)\big)\leq\bar{E}\leq z\pi\gamma T_\text{U}\sum_{k=1}^\infty(k+1)\exp(-\pi\gamma\delta k),
\ee
which follow from the fact that $k\leq\sqrt{k(k+2)}\leq k+1$. This in turn leads to the inequalities
\be
z\pi\gamma T_\text{U}\f{\exp(-2\pi\gamma\delta)}{\big(1-\exp(-\pi\gamma\delta)\big)^2}\leq\bar{E}\leq z\pi\gamma T_\text{U}\f{1}{\big(1-\exp(-\pi\gamma\delta)\big)^2}.
\ee
As a consequence, in order for the mean energy to be larger than a given value, the parameter $\delta=\delta_\beta+\delta_\text{h}$ has to be small. We can write the asymptotic inequalities
\be
z_\text{U}T_\text{U}\f{1+x_-\delta_\beta+y_-\delta_\text{h}}{\pi\gamma\delta^2}+\mathcal{O}(1)\leq\bar{E}\leq z_\text{U}T_\text{U}\f{1+x_+\delta_\beta+y_+\delta_\text{h}}{\pi \gamma\delta^2}+\mathcal{O}(1),
\ee
where $z_\text{U}=\exp(\beta_\text{U}\mu)$ is the value of the fugacity at the Unruh temperature, and $(x_\pm,y_\pm)$ are constant whose explicit form is not necessary. This implies immediately that there is a pair $(x,y)$ of real parameters such that
\be\label{meanE maxwell}
\bar{E}=z_\text{U}T_\text{U}\f{1+x\delta_\beta+y\delta_\text{h}}{\pi\gamma\delta^2}+\mathcal{O}(1).
\ee
Therefore, we obtain that $\bar{E}$ and $\bar{a}_\text{H}=8\pi l\bar{E}$ scale as $\delta_\text{h}^{-2}$ when $\beta=\beta_\text{U}$. This means that the mean horizon area becomes large only if the degeneracy saturates the holographic bound, i.e. if $\lambda=1/4$ in \eqref{degeneracy}.

In order to go further and to derive the entropy, we need to compute the mean number of particles and the partition function itself. These two quantities are simply related by
\be
\bar{n}=z\f{\partial}{\partial z}\log\mathcal{Z}_\text{M}(\beta,\mu)=z\mathcal{Q}(\beta)=z\sum_{k=1}^\infty\exp\left(-\pi\gamma\delta\sqrt{k(k+2)}\right).
\ee
The same strategy as above then leads to the asymptotic expansion
\be
\bar{n}=\log\mathcal{Z}_\text{M}(\beta,\mu)=\f{z_\text{U}}{\pi\gamma\delta}+\mathcal{O}(1).
\ee
As a consequence, the grand canonical entropy $S_\text{M}=\beta(\bar{E}-\mu\bar{n})+\log\mathcal{Z}_\text{M}(\beta,\mu)$ evaluated at the Unruh temperature simplifies at the semi-classical limit ($\delta_\text{h}\ll1$ and $\delta_\beta\ll1$) and takes the form
\be
S_\text{M}=\f{\bar{a}_\text{H}}{4\lp^2}+\f{z_\text{U}}{\pi\gamma\delta}\left(\delta_\beta\left[\f{1}{\delta}-\beta_\text{U}\mu\right]+1-\beta_\text{U}\mu\right)+\mathcal{O}(1),
\ee
where $\mathcal{O}(1)$ is defined with respect to both variables $\delta_\beta$ and $\delta_\text{h}$. The explicit values of $x$ and $y$ introduced in \eqref{meanE maxwell} are not relevant for this calculation. As a consequence we get that
\be
S_\text{M}=\f{\bar{a}_\text{H}}{4\lp^2}+\f{z_\text{U}}{\pi\gamma\delta_\text{h}}(1-\beta_\text{U}\mu)+\mathcal{O}(1),\qquad\text{when}\qquad\delta_\beta=0.
\ee
The entropy of the black hole agrees at leading order with the Bekenstein--Hawking formula but, since $\bar{a}_\text{H}\sim4\pi\gamma\lp^2/\delta^2$, the subleading corrections are generically of order $\sqrt{\bar{a}_\text{H}}$. These corrections are too large compared to the expected logarithmic corrections. However, one can see that for the particular value $\mu=T_\text{U}$ of the chemical potential this contribution vanishes.

In this thermodynamical system, there exists another way of attaining the semi-classical limit. It consists in assuming from the onset that the degeneracy \eqref{degeneracy} saturates the holographic bound, i.e. that $\lambda=1/4$. In that case, the semi-classical limit is obtained only when $\delta_\beta\rightarrow0$, which means that $\beta$ approaches the inverse Unruh temperature. This was studied in \cite{Achour:2014eqa}, where it was shown that the asymptotic expansion of the entropy takes the form
\be
S_\text{M}=\f{\bar{a}_\text{H}}{4\lp^2}+\f{z_\text{U}}{\pi\gamma\delta_\beta}(2-\beta_\text{U}\mu)+\mathcal{O}(1),\qquad\text{when}\qquad\delta_\text{h}=0.
\ee
One can see that in this case the subleading corrections can be eliminated by choosing $\mu=2T_\text{U}$.

In summary, one has that in both cases above a fine tuning of the chemical potential can eliminate the too large subleading corrections to the entropy, which in turns leaves room for hypothetical logarithmic corrections. This expectation is realized in \cite{Achour:2014eqa}, where it has been shown that taking into account polynomial corrections
to the degeneracy (which come from the asymptotic expansion of the number of black holes microstates in LQG) leads to the appearance of logarithmic corrections.

\section{Bose--Einstein statistics and condensation}
\label{sec:3}

\noindent The goal of this section is to reproduce the statistical physics analysis outlined above, but now assuming that the punctures satisfy the Bose--Einstein statistics. We are going to show that, under certain conditions, the presence of a non-vanishing chemical potential leads to the elimination of the too large subleading corrections to the
entropy. The analysis is however more involved than in the case of punctures satisfying the Maxwell--Boltzmann statistics. Furthermore, we will see that at the semi-classical limit the corrections to the entropy are indeed logarithmic only when a condensation phenomenon appears, in the sense that the number $n_{1/2}$ of punctures carrying
a spin $1/2$ becomes much larger to the number of ``excited" (i.e. with spin $j>1/2$) punctures.

First, in order to simplify the analysis, let us assume that the degeneracy \eqref{degeneracy} saturates the holographic bound. This means that $\delta_\text{h}=0$. The case $\delta_\text{h}\neq0$ will be treated later on, in subsection \ref{subsec:holographic mu}. When $\delta_\text{h}=0$, our starting point is the computation of the grand canonical partition function $\mathcal{Z}_\text{B}(\beta,\mu)$ which, following \cite{Ghosh:2013iwa}, is defined by the expression
\be\label{grand canonical partition}
\mathcal{Z}_\text{B}(\beta,\mu)=\prod_j\mathcal{Z}_j(\beta,\mu),\qquad\text{with}\qquad\mathcal{Z}_j(\beta,\mu)=\big(1-z\exp(\beta_\text{U}E_j)\exp(-\beta E_j)\big)^{-1}.
\ee
The product runs once again over the set of half-integers. There is an important difference between this black hole partition function and usual bosonic partition functions of quantum physics. The difference lies in the ``degeneracy" term $\exp(\beta_\text{U}E_j)$, which appears here in the denominator whereas ``degeneracies" typically appear in the nominator in standard quantum systems. The presence of this term here can be traced back to the holographic hypothesis, which postulates the form \eqref{degeneracy} for the degeneracy of quantum microstates. One consequence of this fact is that the partition function is defined only when the following condition is satisfied:
\be\label{condition1}
(\beta-\beta_\text{U})E_j-\beta\mu>0,\qquad\forall\,j\in\mathbb{N}/2.
\ee
The mean energy and the mean number of punctures are respectively given by
\be\label{mean values}
\bar{E}=\sum_jE_j\bar{n}_j\qquad\text{and}\qquad
\bar{n}=\sum_j\bar{n}_j,\qquad\text{with}\qquad\bar{n}_j=\big(z^{-1}\exp\big((\beta-\beta_\text{U})E_j\big)-1\big)^{-1},
\ee
and where $\bar{n}_j$ represents the mean number of punctures colored with the spin representation label $j$. The condition \eqref{condition1} ensures that $\bar{n}_j$ is always positive and thus well-defined. In what follows, it will be useful to decompose the mean number of punctures as $\bar{n}=\bar{n}_{1/2}+\bar{n}_\text{ex}$, where the number of excited punctures is given by
\be
\bar{n}_\text{ex}=\sum_{j\geq1}\bar{n}_j.
\ee
The same decomposition can be done with the mean energy in the form $\bar{E}=\bar{E}_{1/2}+\bar{E}_\text{ex}$, with
\be
\bar{E}_{1/2}=E_{1/2}\bar{n}_{1/2},\qquad\text{and}\qquad\bar{E}_\text{ex}=\sum_{j\geq1}E_j\bar{n}_j.
\ee

We now want to tackle the study of the thermodynamical properties (in particular the entropy) of the system of punctures at the semi-classical limit. If we require only that the mean energy $\bar{E}$ (or equivalently the mean horizon area $\bar{a}_\text{H}$) become large in Planck units at the semi-classical limit, then this semi-classical limit is ill-defined. The reason for this is that, in the grand canonical ensemble, the system admits two intensive free parameters, which are the (inverse) temperature $\beta$ and the chemical potential $\mu$. Indeed, one can achieve the ``large" energy condition by tuning independently only one out of these two parameters. Therefore, we need an extra condition in order to define more precisely the semi-classical limit.

In the first subsection, \ref{subsec:mu zero}, we are going to impose that the temperature is fixed to $\beta_\text{U}$ at the semi-classical limit. Physically, this condition is easily understood from the point of view of quantum field theory in curved space-time. We will show that for the system to become semi-classical, its chemical potential must approach zero. This ensures that the punctures behave as photons at the semi-classical limit and that their number is not fixed.

In the second subsection, \ref{subsec:mu non zero}, we are then going to assume that $\mu$ is fixed (at least in the semi-classical limit) to a non-vanishing (fundamental) value. In this case, we can show that $\beta\rightarrow\beta_\text{c}$ at the semi-classical limit, where $\beta_\text{c}$ is a priori different from the Unruh temperature. However, when $\mu$ is ``small", then $\beta_\text{c}$ approaches $\beta_\text{U}$.

The outcome of these two detailed computations will be the entropy $S_\text{B}$ (where the subscript $\text{B}$ refers to the Bose--Einstein statistics). We are going to show that the leading order term of $S_\text{B}$ reproduces as expected the Bekenstein--Hawking formula, but that the subleading corrections depend specifically on the choice of semi-classical regime. The corrections turn out to logarithmic only when the black hole exhibits a condensation phenomenon, i.e. when the spin $1/2$ representations are dominate in the sense that $\bar{n}_{1/2}\gg\bar{n}_\text{ex}$.

\subsection{First semi-classical regime, $\boldsymbol{\mu\rightarrow0}$ and $\boldsymbol{\beta\rightarrow\beta_{\text{U}}}$}
\label{subsec:mu zero}

\noindent Standard Bose--Einstein condensation (i.e. for usual systems) occurs for a gas of particles when one controls the total number $n$ of particles and the (inverse) temperature $\beta$. There are only two free parameters in the theory. The total energy and the fugacity (or equivalently the chemical potential) ``adapt'' themselves to a situation in which $n$ and $\beta$ are fixed. For instance, the fugacity is fixed by the equation $n=\bar{n}$, where $\bar{n}$ is the mean number of particles. In this section, we assume that the gas of punctures describing the black hole is defined at first sight in a similar way, i.e. that there exists a virtual physical process where one controls the number of punctures and the temperature of the black hole. From this point of view, \eqref{condition1} should be interpreted as a condition on the chemical potential of the system, i.e.
\be
\mu<\left(1-\f{\beta_\text{U}}{\beta}\right)E_j,\qquad\forall\,j\in\mathbb{N}/2.
\ee
This condition takes a very different form depending on wether we are above or below the Unruh temperature:
\bas
&&\text{if}\qquad\beta>\beta_\text{U},\qquad\text{then}\qquad\mu<\left(1-\f{\beta_\text{U}}{\beta}\right)E_{1/2},\\
&&\text{if}\qquad\beta<\beta_\text{U},\qquad\text{then}\qquad\mu=-\infty.
\eas
From this point of view, the value $\beta=\beta_\text{U}$ appears to be very peculiar, and marks a certain transition. At high temperature ($\beta < \beta_\text{U}$), the system necessarily has a negative infinite chemical potential, and therefore it could in principle be described by a classical Maxwell--Boltzmann statistics\footnote{However, the ``classical" Maxwell--Boltzmann partition function \eqref{partition function maxwell} is only formally defined, and one should regularize properly the condition $\mu=-\infty$ in order to make it mathematically well-defined. There exists a simple and natural way to do so, which consists in first truncating the sum over $j$ in the canonical partition function $\mathcal{Q}(\beta)$ to a maximal value $j_\text{max}$, then fixing the chemical potential to an arbitrary value $\mu<(1-\beta_\text{U}/\beta)E_{j_\text{max}}$, and finally taking the limit where $j_\text{max}\rightarrow\infty$.}. At low temperature, the system is well-described by a quantum statistics. Therefore, a quantum-to-classical transition seems to occur at $\beta=\beta_\text{U}$. In what follows, we are going to focus on the ``quantum'' (or low temperature) regime $\beta>\beta_\text{U}$.

In what follows, we are first going to give a brief characterization of the low temperature regime and of the associated physical properties, and then we are going to derive these results more precisely.

\subsubsection{Semi-classical limit and Bose--Einstein condensation}

\noindent As we said in the introduction to this section, we are interested in situations in which the mean energy (or equivalently the mean horizon area) becomes macroscopic (in Planck units). Indeed, this is the minimal requirement in order for the system to be semi-classical. For the time being and for the sake of generality, let us impose no specific conditions on the temperature other than $\beta>\beta_\text{U}$. From the expression \eqref{mean values}, it is immediate to see that the mean energy becomes large at a fixed temperature only when the chemical potential approaches a particular value given by the condition $\epsilon=0$, where the quantity $\epsilon$ (which has the dimension of an energy) is defined by
\be\label{condition on mu}
\epsilon=\left(1-\f{\beta_\text{U}}{\beta}\right)E_{1/2}-\mu.
\ee
More precisely, the straightforward inequality
\be
\bar{E}>\bar{E}_{1/2}=\f{E_{1/2}}{\exp(\beta\epsilon)-1}
\ee
ensures that the horizon area can be arbitrarily large provided that $\epsilon$ is sufficiently small. On the contrary, if $\epsilon$ is finite, one can see that the mean energy is bounded from above, and therefore that the black hole cannot become semi-classical. In fact, when $\beta\neq\beta_\text{U}$, one recovers the usual Bose--Einstein condensation phenomenon which occurs at small temperatures (with respect to the Unruh temperature). Indeed, it is easy to see that the mean number of punctures carrying a spin label $1/2$ is given by
\be
\bar{n}_{1/2}=\f{1}{\exp(\beta\epsilon)-1},
\ee
which is not bounded from above and can take any arbitrarily large value provided that the chemical potential is such that\footnote{When $\beta_\text{U}=0$ (i.e. for usual statistical systems), this condition reduces to the well-known condition that the chemical potential $\mu$ must tend to the energy $E_{1/2}$ of the ground state.} $\epsilon\rightarrow0$. In contrast to this, the mean number $\bar{n}_\text{ex}$ of excited punctures is necessarily bounded according to
\be\label{series Nex}
\bar{n}_\text{ex}\leq\bar{n}_\text{ex}^\text{max},\qquad\text{where}\qquad\bar{n}_\text{ex}^\text{max}=\sum_{j\geq1}\f{1}{\exp\big((\beta-\beta_\text{U})(E_j-E_{1/2})\big)-1}.
\ee
Here we have assumed that we are below the Unruh temperature. Exactly as in Bose--Einstein condensation, when the total number $\bar{n}$ of punctures increases at a given fixed temperature and exceeds the maximal value $\bar{n}_\text{ex}^\text{max}$ for $\bar{n}_\text{ex}$, the punctures ``condensate'' in the spin $1/2$ representation. Therefore, the number of punctures carrying a spin $1/2$ becomes macroscopic.

\subsubsection{Low temperature regime, $\beta\gg\beta_\text{U}$}

\noindent In order to make the previous observation more precise, it would be interesting to obtain a simple expression for $\bar{n}_\text{ex}^\text{max}$ in terms of $\beta$. Unfortunately, this is not possible in general because the series \eqref{series Nex} cannot be written in a simple closed form. Nevertheless, it becomes possible to simplify the expression for $\bar{n}_\text{ex}$ if we assume for instance that the temperature is way below the Unruh temperature, i.e. that $\beta\gg\beta_\text{U}$. This hypothesis is furthermore consistent with the fact that Bose--Einstein condensation occurs experimentally at ``low" temperatures. In this case, it is easy to show that
\be
\text{when}\qquad\beta\gg\beta_\text{U},\qquad\bar{n}_\text{ex}^\text{max}\simeq\exp\big(-\beta(E_1-E_{1/2})\big),\qquad\text{and}\qquad\bar{n}_{1/2}\simeq\f{1}{\beta(E_{1/2}-\mu)}.
\ee
Therefore, we can conclude that at very low temperatures there are almost no excited punctures, and that almost all of the punctures are colored with representations of spin $1/2$. Concerning the chemical potential, its values approaches the minimal energy $E_{1/2}$ at the semi-classical limit according to
\be
\mu\simeq E_{1/2}-\f{1}{\beta\bar{n}}.
\ee
However, the low temperature regime is not the one we are interested in because we would like the black hole temperature at the semi-classical limit (i.e. when the energy is large) to be given by the Unruh temperature and not by $T=0$.

This is the focus of the next subsection, which is devoted to the study of the properties of the system close to the Unruh temperature.

\subsubsection{At Unruh temperature, $\delta_\beta\ll\beta_\text{U}$}
\label{asymptotics}

\noindent Now, we assume that the temperature is very close to the Unruh temperature and we introduce, as in \cite{Ghosh:2013iwa}, the parameter $\delta_\beta=\beta/\beta_\text{U}-1$ (so we have $\delta_\beta\ll\beta_\text{U}$). Furthermore, we impose the condition $\epsilon\ll1$ which, in the case in which $\delta_\beta\ll\beta_\text{U}$, is equivalent to $\mu \ll 1$ for the black hole at the semi-classical limit. Therefore, we study the case $\mu\rightarrow0$ and $\beta\rightarrow\beta_\text{U}$. As we said, the limit $\mu\rightarrow0$ is physically interpreted by the fact that systems of punctures (here quantum excitations of the gravitational field) behave in a way analogous to systems of photons when the chemical potential vanishes \cite{Ghosh:2013iwa}.

In addition to being physically relevant, the regime in which $\epsilon\rightarrow0$ and $\delta_\beta\rightarrow0$ (which implies $\mu\rightarrow0$) is also technically interesting because it allows for explicit calculations (for asymptotic quantities). In this regime, the mean number of particles in the representation of spin $1/2$ is necessarily large, and its asymptotic expansion is given by
\be\label{meanN12}
\bar{n}_{1/2}=\f{1}{\exp(\beta\epsilon)-1}\simeq\f{1}{\delta_\epsilon},\qquad\text{with}\qquad\delta_\epsilon=\beta_\text{U}\epsilon.
\ee
Contrary to what happens in usual Bose--Einstein condensation, the number of excited punctures is also large when $\delta_\beta$ approaches zero. This can be seen on the expression
\ba
\bar{n}_\text{ex}&=&\sum_{j\geq1}\bar{n}_j\nn\\
&=&\sum_{j\geq1}\f{1}{\exp\big(\beta\epsilon+\beta_\text{U}\delta_\beta(E_j-E_{1/2})\big)-1}\nn\\
&=&\sum_{k=2}^\infty\f{1}{\exp\left(\beta\epsilon+\pi\gamma\delta_\beta\big(\sqrt{k(k+2)}-\sqrt{3}\big)\right)-1}\nn\\
&\simeq&\f{1}{\pi \gamma \delta_\beta}\int_{\delta_\beta\Gamma}^\infty\f{\de x}{\exp(\delta_\epsilon)e^x-1},
\ea
where we have introduced $\Gamma=\beta_\text{U}(E_1-E_{1/2})$. The analysis of the last integral leads immediately to the result
\be\label{meanNex}
\bar{n}_\text{ex}\simeq-\f{\log(\Gamma\delta_\beta+\delta_\epsilon)}{\pi\gamma\delta_\beta},
\ee
which shows that the number of excited punctures increases when $\beta$ approaches the Unruh temperature. Therefore, at first sight, it seems that there is no condensation phenomenon at Unruh temperature. However, the situation is not that simple. Indeed, even if $\bar{n}_\text{ex}$ can be arbitrarily large when $\beta$ approaches $\beta_\text{U}$, it can still be negligible when compared to $\bar{n}_{1/2}$. This is the case when
\be\label{BEC condition}
\f{\bar{n}_\text{ex}}{\bar{n}_{1/2}}\simeq-\f{\delta_\epsilon}{\pi\gamma}\f{\log\delta_\beta}{\delta_\beta}\ll1,\qquad\Rightarrow 
\qquad\delta_\epsilon\ll-\f{{\delta_\beta}}{\log\delta_\beta}.
\ee
Conversely, there are situations in which at the semi-classical limit the number of punctures colored with spin $1/2$ is very small. This happens when the condition $\delta_\epsilon\gg-{{\delta_\beta}}/{\log\delta_\beta}$ is satisfied. Therefore, there are different semi-classical regimes depending on the way in which $\beta$ approaches $\beta_\text{U}$ compared to the way in which $\mu$ approaches $0$.

In order to have a more physical meaningful characterization of these conditions, it is useful to rewrite them as conditions involving the mean number $\bar{n}$ of punctures and the mean energy $\bar{E}$. This is in principle possible because, as we said previously, in the presence of a chemical potential the system admits two independent parameters which can be fixed independently and can be chosen freely. To achieve this we need to compute the mean energy, and a straightforward calculation shows that its asymptotic expansion is
\be\label{meanE}
\bar{E}=\bar{E}_{1/2}+\bar{E}_\text{ex},\qquad\text{with}\qquad\bar{E}_{1/2}\simeq E_{1/2}\f{1}{\delta_\epsilon};\qquad\text{and}\qquad\bar{E}_\text{ex}\simeq\f{\pi}{6\gamma\beta_\text{U}}\f{1}{\delta_\beta^2},
\ee
when $\delta_\epsilon\ll1$ and $\delta_\beta\ll1$. Now, we have all the necessary ingredient to classify the different semi-classical regimes. For this, we need to compare $\bar{n}$ given by \eqref{meanN12} and \eqref{meanNex} to $\bar{E}$ given by \eqref{meanN12}. We will assume here that $\gamma$ is a constant of order 1.
\begin{enumerate}
\item There is a semi-classical regime such that $\bar{n}\ll\beta_\text{U}\bar{E}$. For this to be the case, we must necessarily have $\delta_\epsilon\gg\delta_\beta^2$, and therefore the mean energy reduces to
\be
\bar{E}\simeq\f{\pi}{6\gamma\beta_\text{U}}\f{1}{\delta_\beta^2}.
\ee
Then, to extract the leading order term in the expression for $\bar{n}$, we have to further distinguish between the two following subcases:
\bas
\text{(a)}&&\qquad\big(\delta_\beta^2\ll\big)\delta_\epsilon\ll-\f{\delta_\beta}{\log\delta_\beta},\qquad\Rightarrow\qquad\bar{n}\simeq\bar{n}_{1/2}\simeq\f{1}{\delta_\epsilon},\\
\text{(b)}&&\qquad\delta_\epsilon\gg-\f{\delta_\beta}{\log\delta_\beta}\big(\gg\delta_\beta^2\big),\qquad\Rightarrow\qquad\bar{n}\simeq\bar{n}_\text{ex}\simeq-\f{\log(\Gamma \delta_\beta + \delta_\epsilon)}{\gamma \pi\delta_\beta} \,.
\eas
Note that there is a condensation in the first subcase. 
\item There is a semi-classical regime such that $\log\bar{n}\simeq\log(\beta_\text{U}\bar{E})$. For this condition to be satisfied, necessarily $\delta_\epsilon=\mathcal{O}(\delta_\beta^2)$, which implies immediately that
\be
\log(\beta_\text{U}\bar{E})\simeq-\log{\delta_\epsilon},\qquad\text{and}\qquad\bar{n}\simeq\f{1}{\delta_\epsilon}.
\ee
The first condition means that $\bar{E}$ scales as $\delta_\epsilon^{-1}$. Note that there is again a condensation in this semi-classical regime. To be more explicit and to extract precisely the leading order term in the expression of $\bar{E}$, one distinguishes the two different sub cases:
\bas
\text{(a)}&&\qquad\delta_\epsilon\simeq\alpha\delta_\beta^2\qquad\Rightarrow\qquad\bar{E}\simeq\left(E_{1/2}+\f{\alpha\pi}{6\gamma\beta_\text{U}}\right)\f{1}{\delta_\epsilon}\label{E2a},\\
\text{(b)}&&\qquad\delta_\epsilon\ll\delta_\beta^2\hphantom{\alpha}\!\qquad\Rightarrow\qquad\bar{E}\simeq E_{1/2}\f{1}{\delta_\epsilon}.\label{E2b}
\eas
These two regimes are qualitatively the same. 
\item There is no semi-classical limit such that $\bar{n} \gg \beta_\text{U} \bar{E}$. This is clear given the fact that $\bar{E}_j=E_j\bar{n}_j$ for any $j$, with $\beta_\text{U}E_j=2\pi\gamma\sqrt{j(j+1)}$. This implies that there exists a constant $C$ such that $\beta_\text{U}\bar{E}>C\bar{n}$. Therefore, it is not possible to have $\bar{n}\gg\beta_\text{U}\bar{E}$.
\end{enumerate}

Finally, one can say about the chemical potential that it vanishes at the semi-classical limit according to the following behavior:
\be
\mu\simeq E_{1/2}\delta_\beta-T_\text{U}\delta_\epsilon.
\ee
Note that the case $\mu=0$ has been studied in great details in \cite{Ghosh:2013iwa}. With our notations, this situation is equivalent to setting $\delta_\epsilon=\beta_\text{U}E_{1/2}\delta_\beta$ (when $\beta$ is close to $\beta_\text{U}$), which corresponds to the case 1(b). As expected from quantum physics, there is no condensation with zero chemical potential.

\subsubsection{Entropy and corrections to the area law}
\label{asymptotic entropy}

\noindent Now that the different (asymptotic) semi-classical regimes have been properly defined and classified, we can go further into the study of thermodynamical properties of the system. The next step is the computation of the semi-classical entropy $S_\text{B}$. The leading order term is easily obtained, and an immediate calculation shows that
\be
S_\text{B}=\f{\bar{a}_\text{H}}{4\lp^2}+S_\text{cor},\qquad\text{with}\qquad S_\text{cor}=\delta_\beta\beta_\text{U}\bar{E}-\beta\mu\bar{n}+\log\mathcal{Z}_\text{B}(\beta,\mu).
\ee
Now, let us compute $\log\mathcal{Z}_\text{B}(\beta,\mu)$ at the semi-classical limit. It is given by
\ba
\log\mathcal{Z}_\text{B}(\beta,\mu)&=&-\sum_j\log\big(1-\exp\big(\beta\mu+(\beta_\text{U}-\beta)E_j\big)\big)\\
&\simeq&-\log\big(1-\exp(-\delta_\epsilon)\big)-\sum_{j\geq1}\log\big(1-\exp\big(\delta_\beta\beta_\text{U}(E_{1/2}-E_j)-\delta_\epsilon\big)\big).
\ea
Here we have distinguished the spin $1/2$ contribution to the (logarithm of the) partition function from the excited contributions, and we have neglected terms proportional to $\delta_\beta\delta_\epsilon$ in the argument of the last exponential because we are interested only in the leading order terms. An immediate calculation now leads to the asymptotic expression
\ba
\log\mathcal{Z}_\text{B}(\beta,\mu)&\simeq&-\log\delta_\epsilon-\sum_{k=2}^\infty\log\left(1-\exp\left(-\pi\gamma\delta_\beta(\sqrt{k(k+2)}-\sqrt{3})-\delta_\epsilon\right)\right)\\
&\simeq&-\log\delta_\epsilon-\f{1}{\pi\gamma\delta_\beta}\int_{\Gamma\delta_\beta+\delta_\epsilon}^\infty\de x\log\big(1-e^{-x}\big).
\ea
Here we have used the fact that the series can be viewed as a Riemann sum and therefore can be approximated by its corresponding Riemann integral when $\delta_\beta\ll1$. The integral obtained in this way converges, and its main contribution, when the parameters $\delta_\beta$ and $\delta_\epsilon$ are small, is simply given by
\be
\int_{\Gamma\delta_\beta+\delta_\epsilon}^\infty\de x\log\big(1-e^{-x}\big)\simeq\int_{0}^\infty\de x\log\big(1-e^{-x}\big)=-\f{\pi^2}{6}.
\ee
Finally, we end up with the following expression for $\log\mathcal{Z}_\text{B}(\beta,\mu)$ in the semi-classical regime:
\be
\log\mathcal{Z}_\text{B}(\beta,\mu)\simeq-\log\delta_\epsilon+\f{\pi}{6\gamma\delta_\beta}.
\ee

We now have all the necessary ingredients to compute the subleading correction to the entropy as a function of the two independent parameters $\bar{E}$ and $\bar{n}$. There is no simple and explicit formula for $S_\text{cor}$ in general, but from the previous study we can say that the leading order term to $S_\text{cor}$ is necessarily given by the leading order term of the following sum:
\be\label{expansion of S}
\f{\pi}{3\gamma\delta_\beta}+\f{1}{\pi\gamma}\left(\beta_\text{U}E_{1/2}-\f{\delta_\epsilon}{\delta_\beta}\right)\log(\Gamma\delta_\beta+\delta_\epsilon)-\log\delta_\epsilon.
\ee
To write down the leading order term as a function of the macroscopic variables $\bar{E}$ and $\bar{n}$, we need to distinguish between the different cases that were studied in the previous subsection.
\begin{enumerate}
\item Case $\delta_\beta^2\ll\delta_\epsilon$. In this case, $\delta_\beta$ is directly related to the mean energy through
\be
\delta_\beta\simeq\sqrt{\f{\pi}{6\gamma\beta_\text{U}\bar{E}}}.
\ee
The subleading correction to the entropy is then dominated by the first term in \eqref{expansion of S}, which can be written as
\be
S_\text{cor}\simeq\sqrt{\f{2\pi}{3\gamma}\beta_\text{U}\bar{E}}.
\ee
For the sake of completeness, let us give formulae which express the remaining physical parameters $\mu$ and $\bar{n}$ in terms of the small parameters $\delta_\epsilon$ and $\delta_\beta$. In order to do so, we further need to distinguish between different subcases.
\begin{enumerate}
\item When $\delta_\beta^2\ll\delta_\epsilon\ll-\delta_\beta/\log\delta_\beta$, we have
\be
\delta_\epsilon\simeq\f{1}{\bar{n}},\qquad\text{and}\qquad\mu\simeq E_{1/2}\delta_\beta.
\ee
\item When $\delta_\epsilon\gg-\delta_\beta/\log\delta_\beta$, we have again to distinguish between different subcases.
\begin{enumerate}
\item If $\delta_\epsilon\gg\delta_\beta$, then
\be
\delta_\epsilon\simeq\exp\left(-\pi\bar{n}\sqrt{\f{\gamma \pi}{6\beta_\text{U}\bar{E}}}\right),\qquad\text{and}\qquad\mu\simeq-T_\text{U}\delta_\epsilon.
\ee
\item If $\delta_\epsilon\simeq\delta_\beta$, there exists a constant $C\in\mathbb{R}$ such that $\delta_\epsilon\simeq C\delta_\beta$, and then
\be
\delta_\epsilon\simeq C\sqrt{\f{2\pi}{3\gamma\beta_\text{U}\bar{E}}},\qquad\text{and}\qquad\mu\simeq(E_{1/2}-CT_\text{U})\delta_\beta.
\ee
\item If $\delta_\epsilon\ll\delta_\beta$, then the expression for $\delta_\epsilon$ in terms of $\bar{E}$ and $\bar{n}$ is rather complicated and not very useful. For this reason, we will not write it. However, the chemical potential it is given by $\mu\simeq E_{1/2}\delta_\beta$.
\end{enumerate}
\end{enumerate}
\item Case $\delta_\epsilon=\mathcal{O}(\delta_\beta^2)$. We have seen that in this regime we have $\bar{E}\simeq E_0\bar{n}$, where $E_0=E_{1/2}$ when $\delta_\epsilon \ll\delta_\beta^2$ (see \eqref{E2b}), and $E_0=E_{1/2}+\alpha\pi\beta_\text{U}^2/(6\gamma)$ when $\delta_\epsilon\simeq\alpha\delta_\beta^2$ (see \eqref{E2a}). Therefore, we have that
\be
\delta_\epsilon\simeq\f{1}{\bar{n}},\qquad\text{and}\qquad\delta_\beta^2\simeq\f{\pi}{6\gamma}\f{1}{\beta_\text{U}(\bar{E}-E_{1/2}\bar{n})},
\ee
which implies that the subleading corrections are of the form
\be\label{log corrections}
S_\text{cor}\simeq\log(\beta_\text{U}\bar{E})+\f{\pi}{3\gamma}\f{1}{\delta_\beta}.
\ee
\end{enumerate}

The study of all these different asymptotic cases leads to the conclusion that the corrections to the entropy cannot be logarithmic in $\bar{E}$ (and therefore in $\overline{a}_H$) in case $(1)$, where they indeed turn out to be of the form $\mathcal{O}(\sqrt{\bar{E}})$. However, the correction can be logarithmic in case $(2)$ if, in addition, the condition  
$-\delta_\beta\log\delta_\epsilon\gg1$ (implying that $\delta_\epsilon^{-1}\gg\exp(\delta_\beta^{-1})$) is satisfied. In this case, we see that \eqref{log corrections} leads to
\be
S_\text{B}\simeq\f{\bar{a}_\text{H}}{4\lp^2}+\log\bar{a}_\text{H}.
\ee
The physical meaning of the condition $-\delta_\beta \log\delta_\epsilon\gg1$ is not so clear, but at the mathematical level it says that $\delta_\epsilon$ approaches zero much faster (in fact at least exponentially faster) than $\beta$ approaches $\beta_\text{U}$. This happens for instance when $\delta_\beta$ is small (compared to $\beta_\text{U}$) but constant. In this case, when the black hole horizon becomes larger and larger, a Bose--Einstein condensation occurs (at ``high" temperature). When expressed in terms of means values, the previous condition can be written as
\be
\bar{E}\simeq\bar{E}_{1/2}\gg E_{1/2}\exp\sqrt{\f{6\gamma\beta_\text{U}}{\pi}\bar{E}_\text{ex}} .
\ee
Then, asking that the energy of the punctures carrying a spin $1/2$ be exponentially larger than the square root of the excited energy, one gets logarithmic subleading corrections to the entropy.

\subsection{Second semi-classical regime, $\boldsymbol{\mu\neq0}$}
\label{subsec:mu non zero}

\noindent Now, we investigate the case in which the chemical potential is a fundamental constant of the theory (at least at the semi-classical limit), and does therefore not depend on the number of punctures or on the temperature. This is in sharp contrast to what usually happens in Bose--Einstein condensation. There is only one free parameter in the theory. Furthermore, we assume that there exists a physical process (matter collapse for instance) which causes the horizon area (or equivalently the energy of the black hole) to increase and to become macroscopic. The number of punctures and their colors change during this process, and we are going to show how these two quantities behave when the horizon area becomes macroscopic.

As opposed to the case in which $\mu\rightarrow0$, there exists here only one semi-classical regime for which the horizon area becomes macroscopic. We are going to see that this semi-classical regime is also characterized by a condensation of the spin labels to the value $1/2$. However, this condensation occurs at a temperature different from the Unruh temperature.

\subsubsection{Asymptotic expansion of $\bar{n}$  and $\bar{E}$ at the semi-classical limit}

\noindent In order to characterize the semi-classical limit when $\mu$ is fixed, it will be convenient to use the following expression for the mean number of punctures colored by the spin $j$:
\be\label{Phi and delta}
\bar{n}_j=(\exp\Phi_j-1)^{-1},\qquad\text{with}\qquad\Phi_j=\beta_\text{U}(E_j-\mu)\delta_\mu+\mu\beta_\text{U}\f{E_j-E_{1/2}}{E_{1/2}-\mu},
\ee
where we have introduced
\be
\delta_\mu=\f{\beta-\beta_\text{c}}{\beta_\text{U}},\qquad\text{with}\qquad\beta_\text{c}=\f{E_{1/2}}{E_{1/2}-\mu}\beta_\text{U}.
\ee
Note that $\delta_{\mu=0}=\delta_\beta$ (where $\delta_\beta$ was introduced in sections). The system is the only defined when $\Phi_j>0$, which implies necessarily that
\be
\mu<E_{1/2},\qquad\text{and}\qquad\delta_\mu>0.
\ee
Since $\mu$ is now fixed, the condition on $\beta$ is now changed, and the semi-classical limit occurs a priori at $\beta_\text{c}\neq\beta_\text{U}$. However, we will see later on that it is possible to take the limit $\mu\rightarrow0$ in order to fix the inverse temperature to $\beta_\text{U}$ at the semi-classical limit.

Now, notice that we have
\be\label{Phi behavior}
\Phi_{1/2}=\beta_\text{U}(E_{1/2}-\mu)\delta_\mu,\qquad\text{and}\qquad\Phi_{j\geq1}>\mu\beta_\text{U}\f{E_j-E_{1/2}}{E_{1/2}-\mu}.
\ee
This ensures that $\Phi_{1/2}$ can be as close to zero as we want, whereas the quantities $\Phi_j$ have non-zero minima when $j\geq1$. These minima are reached when $\beta$ approaches $\beta_\text{c}$. These properties are the signature of a condensation phenomenon when $\beta$ approaches $\beta_\text{c}$.
Notice that this is a priori no longer true when $\mu=0$, since in this case every $\Phi_j$ can approach zero with no restriction when $\beta$ tends to $\beta_\text{c}$.

The bound \eqref{Phi behavior} on $\Phi_j$ implies immediately that $\bar{n}_\text{ex}$ and $\bar{E}_\text{ex}$ are bounded from above, and cannot exceed the maximal values $N_\text{ex}^\text{max}$ and $E_\text{ex}^\text{max}$ defined by
\ba
&&E_\text{ex}^\text{max}=\f{\gamma\pi}{\beta_\text{U}}\xi\left(\f{\mu\beta_\text{c}}{\sqrt{3}}\right),\qquad\text{with}\qquad\xi(x)=\sum_{k=2}^\infty\f{\sqrt{k(k+2)}}{\exp\left(x\big(\sqrt{k(k+2)}-\sqrt{3}\big)\right)-1},\\
&&N_\text{ex}^\text{max}=\zeta\left(\f{\mu\beta_\text{c}}{\sqrt{3}}\right),\hphantom{\f{\gamma\pi}{\beta_\text{U}}}\qquad\text{with}\qquad\zeta(x)=\sum_{k=2}^\infty\f{1}{\exp\left(x\big(\sqrt{k(k+2)}-\sqrt{3}\big)\right)-1}.
\ea
Since $\mu$ is fixed (to a non-vanishing value), $N_\text{ex}^\text{max}$ and $E_\text{ex}^\text{max}$ depend only on the fundamental constants of the theory.

One consequence of this fact is that when the mean energy $\bar{E}$ (or equivalently the mean horizon area) exceeds the value $E_\text{ex}^\text{max}$, only spin $1/2$ punctures contribute to the increasing energy (or mean horizon area), and we recover the Bose--Einstein condensation. Furthermore, the energy becomes macroscopic only when the temperature approaches the critical temperature, i.e. when $\beta\rightarrow\beta_\text{c}$. In this case, we have the asymptotic expansion
\be
\bar{E}_{1/2}=\f{E_{1/2}}{\beta_\text{U}(E_{1/2}-\mu)}\f{1}{\delta_\mu}+\mathcal{O}(1),
\ee
where $\bar{E}_{1/2}$ is viewed as a function of $\delta_\mu$. As a consequence, when the temperature approaches the critical temperature such that the mean area is large enough, i.e.
\be
\f{\bar{a}_\text{H}}{4\lp^2}>\gamma\pi\xi\left(\f{\mu\beta_\text{c}}{\sqrt{3}}\right),
\ee
then the number of excited punctures is ``saturated'' and any increase in the area of the horizon is due to the addition of punctures carrying a spin $1/2$. In this case, the mean area is related to the temperature as follows:
\be
\f{\bar{a}_\text{H}}{4\lp^2}=\f{\beta_\text{c}}{\beta_\text{U}}\f{1}{\delta_\mu}+\mathcal{O}(1).
\ee 

Outside of this regime, the equation of state is more complicated to obtain explicitely. 
For the same reasons, the asymptotic behavior of the mean number of punctures when $\delta_\mu\rightarrow0$ is given by
\be
\bar{n}=\bar{n}_{1/2}+\mathcal{O}(1),\qquad\text{with}\qquad\bar{n}_{1/2}=\f{1}{\beta_\text{U}(E_{1/2}-\mu)}\f{1}{\delta_\mu}+\mathcal{O}(1)=\f{1}{\gamma\pi\sqrt{3}}\f{\beta_\text{c}}{\beta_\text{U}}\f{1}{\delta_\mu}+\mathcal{O}(1).
\ee
Therefore, at the semi-classical limit, when the mean horizon area $\bar{a}_\text{H}$ becomes macroscopic, the number of punctures carrying spin $1/2$ increases drastically. There is a condensation of spin $1/2$ representations, exactly as there is a Bose--Einstein condensation of particles in the ground state in usual quantum physics. The condition for the condensation to occur is given by
\be\label{condensation condition}
\bar{n}_{1/2}\gg\zeta\left(\f{\mu\beta_\text{c}}{\sqrt{3}}\right).
\ee

\subsubsection{Entropy and corrections to the area law}

\noindent In order to compute the entropy, we need to compute the asymptotic expansion of $\log\mathcal{Z}_\text{B}(\beta,\mu)$ in the limit $\beta\rightarrow\beta_\text{c}$. Using the notations introduced above, the partition function is given by
\be
\log\mathcal{Z}_\text{B}(\beta,\mu)=-\sum_j\log\big(1-\exp(-\Phi_j)\big).
\ee
Again, we study separately the contribution from the punctures with spin $1/2$ and from the excited punctures. The contribution of the punctures carrying spin $1/2$ is given by
\be
-\log\big(1-\exp(-\Phi_{1/2})\big)=-\log\big(1-\exp\big(\beta_\text{U}(\mu-E_{1/2})\delta_\mu\big)\big)=-\log\delta_\mu+\mathcal{O}(1).
\ee
The contribution of the excited punctures is irrelevant for the asymptotic of $\log\mathcal{Z}_\text{B}(\beta,\mu)$ when $\beta$ approaches $\beta_\text{c}$. Indeed, when $\delta_\mu=0$ (i.e. when $\beta=\beta_\text{c}$), contribution from the excited punctures is well-defined (i.e. the series is convergent), as we now show. First, if $\beta=\beta_\text{c}$,
\be
\sum_{j\geq1}\log\big(1-\exp(-\Phi_j)\big)=\sum_{j\geq1}\log\left(1-\exp\left[\mu\beta_\text{c}\left(1-\f{E_j}{E_{1/2}}\right)\right]\right).
\ee
This series is easily proven to be convergent since the summand is equivalent (when $j$ becomes large) to
\be
\log\left(1-\exp\left[\mu\beta_\text{c}\left(1-\f{E_j}{E_{1/2}}\right)\right]\right)\simeq-\exp\left[\mu\beta_\text{c}\left(1-\f{E_j}{E_{1/2}}\right)\right]
\simeq-\exp\left[\mu\beta_\text{c}\left(1-\f{n}{\sqrt{3}}\right)\right],
\ee
whose series is convergent. As a consequence, we obtain the asymptotic expansion
\be
\log\mathcal{Z}_\text{B}(\beta,\mu)=\log\delta_\mu+\mathcal{O}(1).
\ee
The computation of the semi-classical expansion of the entropy $S_\text{B}=\beta(\bar{E}-\mu\bar{n})+\log\mathcal{Z}_\text{B}(\beta,\mu)$ is therefore immediate and leads to
\be
S=\f{\bar{a}_\text{H}}{4\lp^2}+\log\bar{a}_\text{H}+\mathcal{O}(1).
\ee
We recover the Bekenstein--Hawking expression with logarithmic corrections.
 
\subsubsection{The limit $\mu\rightarrow0$}

\noindent To finish with this case, let us finally assume that we start with a finite value of $\mu$, first consider the limit $\delta_\mu\rightarrow0$ (as we did in this section), and then we assume that $\mu\rightarrow0$ together with the condition \eqref{condensation condition}. This condition ensures that a condensation occurs and also that the corrections to the Bekenstein--Hawking entropy are logarithmic.
 
To show that this is indeed the case, it is useful to establish a relationship between the classical regime describe here ($\mu$ fixed to a non-zero value) and the regimes described in the previous subsections. A straightforward calculation shows that the limit $\beta\rightarrow\beta_\text{c}$ corresponds to
\be
\delta_\beta=\delta_\mu+\f{\mu}{E_{1/2}-\mu}\rightarrow\f{\mu}{E_{1/2}-\mu},\qquad\text{and}\qquad T_\text{U}\delta_\epsilon=-\mu+\f{\beta-\beta_\text{U}}{\beta}E_{1/2}\rightarrow0.
 \ee
Therefore, if we first take the limit $\delta_\epsilon\rightarrow0$ with $\mu$ fixed, and then send $\mu$ to zero, we are clearly in case (2), where $\delta_\epsilon\ll\delta_\beta^2$. In this case, the correction to the entropy is indeed logarithmic.
 
\subsection{Holographic bound with a chemical potential}
\label{subsec:holographic mu}

\noindent This last subsection is devoted to the study of the case $\delta_\text{h}\neq0$, i.e. when we do not assume from the beginning that the holographic bound is saturated. The expressions for the grand canonical partition function \eqref{grand canonical partition} and the related thermodynamical quantities \eqref{mean values} are formally the same as above, simply with $\beta_\text{U}$ replaced by $(1-\delta_\text{h})\beta_\text{U}$. Therefore, the system now admits three free parameter: the temperature (or equivalently $\delta_\beta$), the chemical potential $\mu$, and the holographic parameter $\delta_\text{h}$. There is a priori more freedom to reach the semi-classical regime.

\subsubsection{Semi-classical regime and holographic bound}

\noindent An immediate analysis shows that in order for the black hole to become macroscopic, we must impose that the parameter $\epsilon_\text{h}$ (which satisfies $\epsilon_\text{h}>0$), defined by
\be
\epsilon_\text{h}=\f{\beta_\text{U}}{\beta}E_{1/2}\delta-\mu,\qquad\text{with}\qquad\delta=\delta_\text{h}+\delta_\beta,
\ee
approaches zero, i.e. $\epsilon_\text{h}\ll1$. Note that $\epsilon_{\text{h}=0}=\epsilon$, with $\epsilon$ given by \eqref{condition on mu}.

If in addition to this we add the condition that the (inverse) temperature tend to $\beta_\text{U}$, then we get necessarily that $\delta_\beta\ll1$, and the condition $\epsilon_\text{h}\ll1$ becomes
\be
E_{1/2}\delta_\text{h}-\mu\ll1.
\ee
As shown in \cite{Ghosh:2013iwa}, the requirements that the area be large and the temperature be fixed to $\beta_\text{U}$ necessarily imply that the holographic bound is saturated when there is no chemical potential $\mu=0$. We also recover what we have just shown in the previous section, namely that the two previous semi-classical requirements imply that the chemical potential vanishes in the semi-classical regime if we set $\delta_\text{h}=0$ from the beginning.

Here, we see that neither $\delta_\text{h}$ nor $\mu$ are uniquely fixed by the requirement of semi-classicallity, as opposed to what could have been expected from the introduction of a new free parameter in the model. However, in light of the physical reasons discussed in the previous section, an additional natural requirement is to ask that the chemical potential of the black hole be ``small"  ($\beta_\text{U}\mu\ll1$) in the semi-classical regime. This condition implies at the end of the day that the degeneracy bound must be saturated in the semi-classical regime, which can be summarized as follows:
\be
\text{Semi-classical regime}\quad\left(\beta\rightarrow\beta_\text{U},\qquad\mu\rightarrow0,\qquad\f{\bar{a}_\text{H}}{\lp^2}\rightarrow\infty\right),\qquad
\Rightarrow\qquad\delta_\text{h}\rightarrow0.
\ee
Adding the condition $\mu\rightarrow0$ as a requirement for the semi-classical limit makes the system equivalent (in this semi-classical limit) to the one studied in \cite{Ghosh:2013iwa} where $\mu=0$. Therefore, it is not a surprise that we here recover the saturation of the holographic bound.

\subsubsection{Entropy and corrections to the area law}

\noindent The study of the asymptotic expansion of the mean numbers of punctures ($\bar{n}_{1/2}$ and $\bar{n}_\text{ex}$) and the mean energies ($\bar{E}_{1/2}$ and $\bar{E}_\text{ex}$) is exactly the same as in subsection \eqref{asymptotics}. The only difference here is that we simply have to replace the small parameter $\delta_\beta$ by the small parameter $\delta=\delta_\beta+\delta_\text{h}$. However, the analysis of the semi-classical entropy and of the subleading corrections could differ.

Following exactly the same analysis as that of the previous subsection, we have
\be\label{S with h}
S_\text{cor}\simeq\beta_\text{U}(\delta_\beta\bar{E}-\mu\bar{n})+\log\mathcal{Z}_\text{B}(\beta,z),
\ee
whose leading order term is necessarily the leading order term of the sum
\be
\left(1+\f{\delta_\beta}{\delta}\right)\f{\pi}{6\gamma \delta}-\beta_\text{U}E_{1/2}\f{\delta_\text{h}}{\delta_\epsilon}+\f{1}{\pi\gamma}\left(\beta_\text{U}E_{1/2}-\f{\delta_\epsilon}{\delta}\right)\log(\Gamma\delta+\delta_\epsilon)-\log\delta_\epsilon .
\ee
Note that, at the difference with the mean energy and the mean number of punctures, the entropy is not symmetric under the exchange $\delta_\beta\leftrightarrow\delta_\text{h}$ because of the first term in \eqref{S with h}. For this reason, the behavior of the entropy at the semi-classical limit is different from what was studied in the previous subsection.

When $\delta_\text{h}=0$, we recover the asymptotic expansion \eqref{expansion of S}. If we assume that $\delta_\text{h}$ is ``small enough" in comparison to $\delta_\beta$ and $\delta_\epsilon$, the analysis of subsection \eqref{asymptotic entropy} still holds as well. The general case, in which $\delta_\text{h}=\mathcal{O}(\delta_\beta)$ (which means that $\delta_\beta$ does not tend to zero faster than $\delta_\text{h}$), is more subtle to study. Since it is not physically relevant for the present study, we will not perform its analysis here.

Here we are more interested in the case $\delta_\beta=0$, which means that we fix the temperature to be the Unruh temperature (as in \cite{Ghosh:2013iwa}). The novelty compared to 
the previous subsection is the presence of a term proportional to $\delta_\text{h}/\delta_\epsilon$ in the asymptotic expansion \eqref{S with h}. This implies that for the subleading corrections to the entropy to be logarithmic it is necessary to satisfy the following requirements:
\begin{enumerate}
\item The term $-\log\delta_\epsilon$ dominate the asymptotic expansion, which implies in particular that 
\be
-\delta_\epsilon\log\delta_\epsilon\gg\delta_\text{h}.
\ee
\item The condition $\delta_\epsilon\ll\delta_\text{h}^2$ must hold in order to have $\delta_\epsilon\propto1/\bar{E}$ (i.e. we are in the case (2) studied above).
\end{enumerate}

These conditions are clearly contradictory. Therefore, the entropy cannot a priori have logarithmic corrections if we impose $\beta=\beta_\text{U}$ and $\delta_\text{h}\neq0$, even if $\delta_\text{h}$ is arbitrary small.

\section{Conclusion}
\label{sec:4}

\noindent In this paper, we have studied the thermodynamical properties of a black hole viewed as a system of non-interacting punctures with an a priori non-vanishing chemical potential. We have started by assuming that the punctures satisfy the classical Maxwell--Boltzmann statistics. With this choice of statistics, we have shown that a fine tuning of the chemical potential (to a non-vanishing value) leads to logarithmic corrections to the entropy, and eliminates the large $\sqrt{a_\text{H}}$ corrections found previously in \cite{Ghosh:2013iwa} when $\mu$ was fixed to zero from the beginning. On the one hand, this result is appealing because it allows to reproduce the expected corrections for the entropy, but on the other hand it requires a non-zero value for the chemical potential, which should be given a physical explanation. The value of $\mu$ is of the order of $T_\text{U}$, and could therefore be related to the fact that the black hole is in equilibrium with the emitted radiation. So far, we cannot propose any physical interpretation for this result, which therefore should be viewed only as a toy model where it is shown explicitly that the presence of a chemical potential can change the subleading corrections to the entropy.
 
Then, we have made the assumption that the punctures satisfy the Bose--Einstein statistics, and considered also an a priori non-vanishing chemical potential. We have studied the semi-classical properties of the system in great details, and have shown that there exist different inequivalent semi-classical regimes. From the physical point of view, there is a particularly interesting regime in which the corrections to the entropy are logarithmic (as in the case of the Maxwell--Boltzmann statistics). Furthermore, the logarithmic corrections come together with a Bose--Einstein condensation of the punctures, in the sense that the system is dominated by punctures carrying the minimal (or ground state) spin value $1/2$. The idea that a Bose--Einstein condensation could occur for black holes in LQG is very appealing and is clearly compatible with the usual result stating that small spins dominate at the semi-classical limit. However, this requires the hypothesis of a holographic degeneracy to be valid all the way down to the deep Planckian regime. This can unfortunately no longer be justified by the standard quantum field theoretical argument used in \cite{Ghosh:2013iwa}, which is an argument coming from low energy effective field theory.

To summarize, we have seen that one can recover logarithmic corrections in a non-trivial way when a chemical potential is present, even if at the semi-classical limit this chemical potential can eventually vanish (which is the case with the Bose--Einstein statistics but not with the Maxwell--Boltzmann statistics). However, when the chemical potential is chosen to be vanishing from the onset the corrections are larger than logarithmic corrections. Therefore, the presence of a chemical potential seems to be related to the logarithmic corrections in this description of quantum black holes as gases of punctures, and it would be interesting to study and understand this relationship further. A possible interesting generalization of the present work would be to investigate the case in which the area spectrum is continuous, which appears in the context of the analytic continuation of the black hole entropy calculation \cite{FGNP,Achour:2014eqa}.

\end{document}